\documentclass[aps,prl,twocolumn,showpacs,groupedaddress]{revtex4}
\pdfoutput=1
\usepackage{graphicx}
\usepackage{amsmath}
\usepackage{mathrsfs}
\usepackage{amsmath}
\usepackage{amsfonts}
\usepackage{amssymb}
\usepackage{epstopdf}

\newcommand{\ket}[1]{{\left| {#1} \right\rangle}}

\begin{document}
\title{Modular Entanglement of Atomic Qubits using both Photons and Phonons}
\author{D. Hucul}\email[dhucul@umd.edu]{}
\author{I.~V. Inlek}
\author{G. Vittorini}
\author{C. Crocker}
\author{S. Debnath}
\author{S.~M. Clark}
\altaffiliation{Present Address: Sandia National Laboratories, Albuquerque, NM, USA}
\author{C. Monroe}
\affiliation{Joint Quantum Institute, University of Maryland Department of Physics and  
                  National Institute of Standards and Technology, College Park, MD  20742 }
\date{\today}
\maketitle

%\linenumbers
\textbf{Quantum entanglement is the central resource behind applications in quantum information science, from quantum computers \cite{MikeAndIke} and simulators of complex quantum systems \cite{NatPhysQSIM} to metrology \cite{Metrology} and secure communication \cite{MikeAndIke}.  All of these applications require the quantum control of large networks of quantum bits (qubits) to realize gains and speedups over conventional devices. However, propagating quantum entanglement generally becomes difficult or impossible as the system grows in size, owing to the inevitable decoherence from the complexity of connections between the qubits and increased couplings to the environment.  Here, we demonstrate the first step in a modular approach \cite{musiqcpaper} to scaling entanglement by utilizing a hierarchy of quantum buses on a collection of three atomic ion qubits stored in two remote ion trap modules.  Entanglement within a module is achieved with deterministic near-field interactions through phonons \cite{blatt:2008}, and remote entanglement between modules is achieved  through a probabilistic interaction through photons \cite{DuanRMP}.  This minimal system allows us to address generic issues in synchronization and scalability of entanglement with multiple buses, while pointing the way toward a modular large-scale quantum computer architecture that promises less spectral crowding and less decoherence.  We generate this modular entanglement faster than the observed qubit decoherence rate, thus the system can be scaled to much larger dimensions by adding more modules.}

Small modules of qubits have been entangled through native local interactions in many physical platforms, such as trapped atomic ions through their Coulomb interaction \cite{blatt:2008}, Rydberg atoms through their electric dipoles \cite{Saffman2010, Grangier2010}, nitrogen-vacancy centers in diamond through their magnetic dipoles \cite{dolde:2013}, and superconducting Josephson junctions through capacitive or inductive couplings \cite{neeley:2010, dicarlo:2010}.  However, each of these systems is confronted with practical limits to the number of qubits that can be reliably controlled, stemming from inhomogeneities, the complexity and density of the interactions between the qubits, or quantum decoherence.  Scaling beyond these limits can be achieved by invoking a second type of interaction that can extend the entanglement to other similar qubit modules.  Such an architecture should therefore exploit both the local interactions within the qubit modules, and also remote interactions between modules (an example architecture is shown in Fig 1). Optical interfaces provide ideal buses for this purpose \cite{CZKM97,DLCZ}, as optical photons can propagate over macroscopic distances with negligible loss. Several qubit systems have been entangled through remote optical buses, such as atomic ions \cite{moehring:2007}, neutral atoms \cite{nolleke:2013}, and nitrogen-vacancy centers in diamond \cite{bernien:2013}.  

\begin{figure*}
\includegraphics[width=6.5 in]{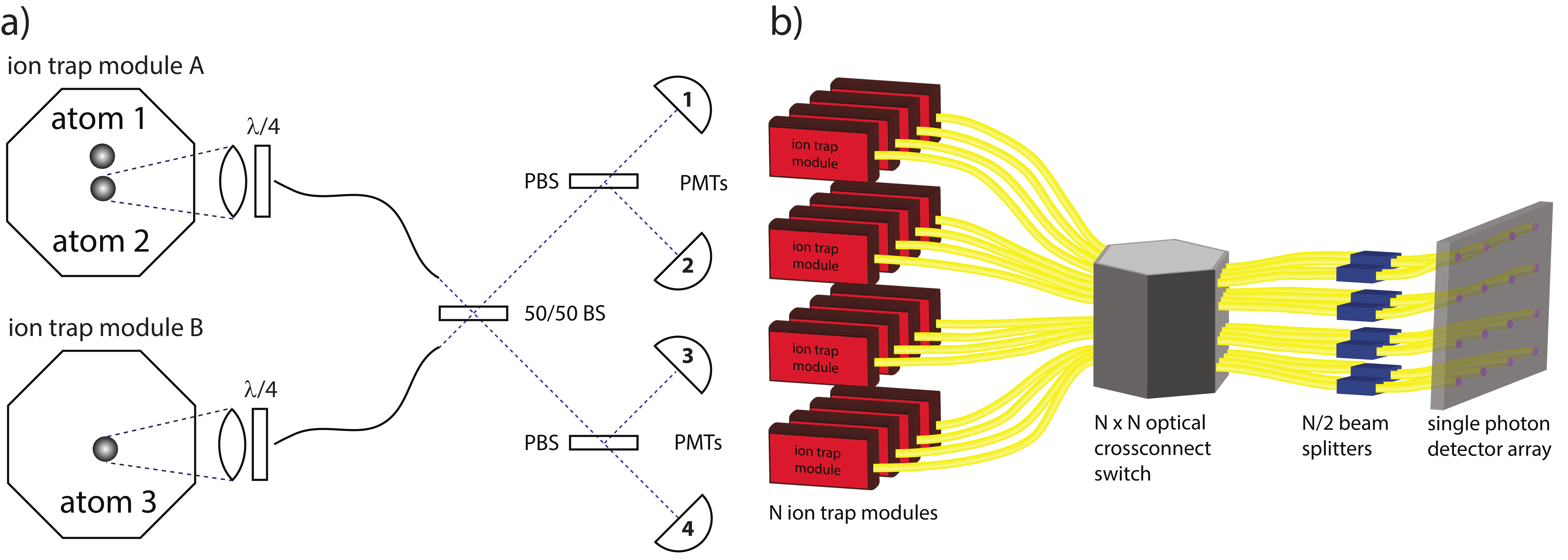}
\caption{\textbf{a)} Experimental setup. Two modules separated by $\sim$1 meter each contain an ion trap. High numerical aperture objectives couple spontaneously emitted photons from a single atom into a single-mode optical fiber. The photons from atoms in separate traps interfere on a 50/50 beam-splitter, are sorted by polarizing beam-splitters then detected by photomultiplier tubes (PMTs). Coincident detection of photons on specific PMT pairs heralds entanglement of atomic spins (see Methods Summary). \textbf{b)} Schematic of a large-scale, modular quantum network of trapped ions. Ion trap modules (red boxes) confine atoms coupled together through their Coulomb bus, and entanglement within modules is accomplished with the application of spin dependent forces to the trapped atoms [4]. Probabilistic, heralded entanglement is generated between modules via interference of emitted photons from each module. A reconfigurable N x N cross connect switch links arbitrary modules. Photon interference occurs at fiber beam-splitters, and a single photon detector array heralds entanglement of atomic spins between modules.}
\label{fig:network and dream}
\end{figure*}

In the experiment reported here, we juxtapose local and remote entanglement buses utilizing trapped atomic ion qubits, balancing the requirements of each interface within the same qubit system. The observed entanglement rate within and between modules is faster than the observed entangled qubit decoherence rate.  Surpassing this threshold implies that this architecture can be scaled to much larger systems, where entanglement is generated faster than coherence is lost.

The qubits in this experiment are defined by the two hyperfine `clock' states, $\ket{F = 0, m_F = 0} \equiv \ket{0}$ and  $\ket{F = 1, m_F = 0} \equiv \ket{1}$, which are separated by $\omega_0 = 2\pi \times 12.64282$ GHz in the ${}^{2}S_{1/2}$ manifold of trapped ${}^{171}$Yb$^{+}$ atoms. Laser cooling, optical pumping, and readout occur via standard state-dependent fluorescence techniques \cite{olmschenk:2007}. The qubits are trapped in two independent modules separated by $\sim$1 meter as shown in Fig. \ref{fig:network and dream}a. (The ion traps, light collection optics, and interferometer could in principle be part of a modular, scalable architecture as shown in Fig. \ref{fig:network and dream}b.)

\begin{figure*}[tbp]
\includegraphics[width=4 in]{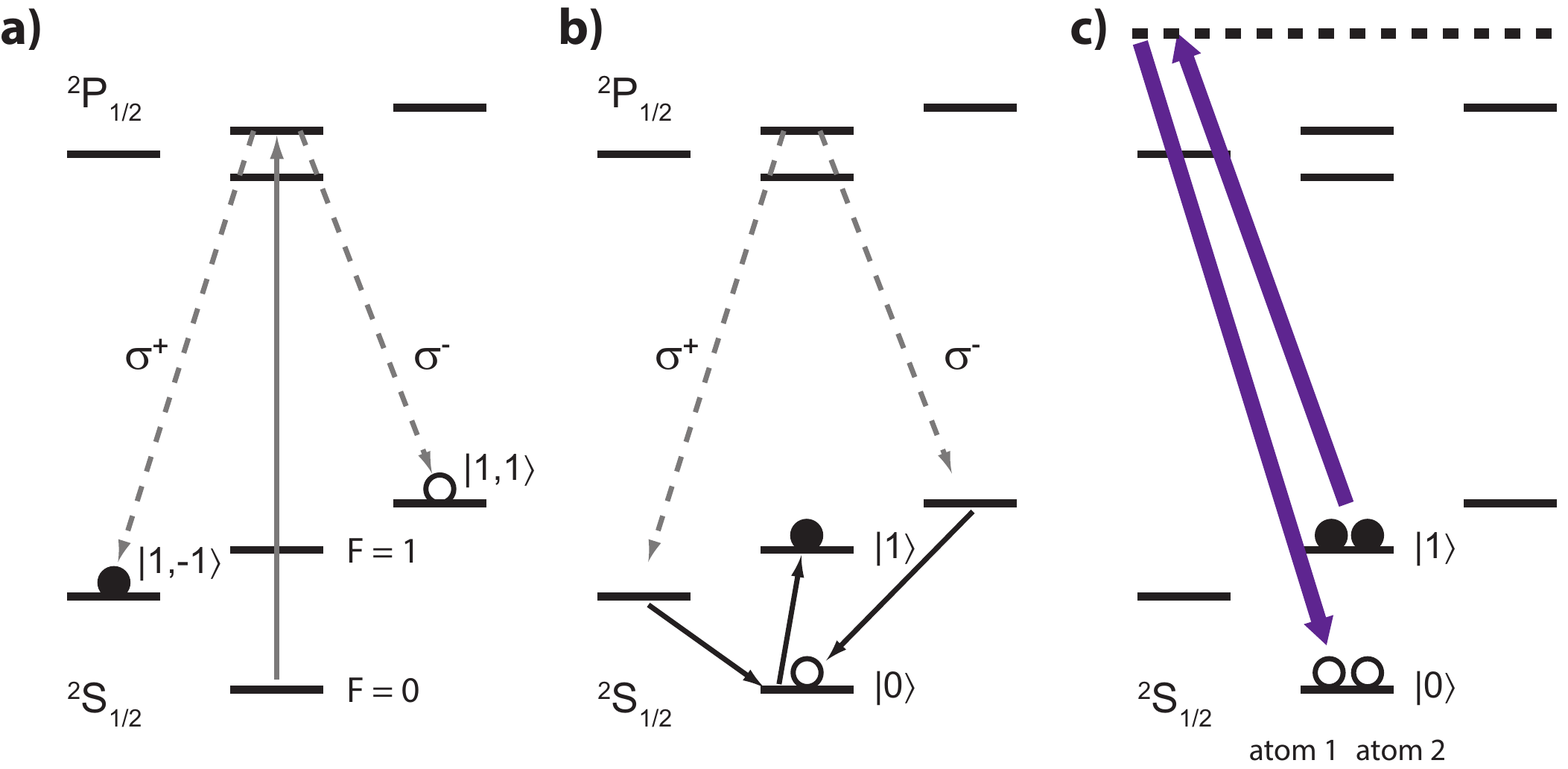}
\caption{\textbf{a)} Resonant excitation scheme and single photon emission in Yb${}^+$ atom system. After optically pumping the atoms to the $\ket{F,m_F} = \ket{0,0}$ state of the ${}^2S_{1/2}$ manifold, a frequency-doubled, mode-locked Ti:sapphire laser excites the atom to the $\ket{1,0}$ state of the ${}^2P_{1/2}$ manifold whereby the atom decays to the $\ket{1,\pm 1}$ states via emission of $\sigma^{\mp}$ polarized photons into optical fibers. \textbf{b)}.  After interference of the two photons on a 50/50 non-polarizing beam-splitter, we apply a series of microwave transfer pulses to transfer the entangled state to the clock basis, resulting in the state 
$\ket{01} + e^{i\phi_{AB}} \ket{10}$ where $\phi_{AB}$ is the intermodular phase. \textbf{c)} We entangle atomic spins within module A through spin dependent optical dipole forces [5, 23] .}
\label{fig:network and microwaves}
\end{figure*}

In order to generate remote entanglement between atoms in physically separated ion trap modules, we synchronously excite each atom with a resonant fast laser pulse \cite{moehring:2007}.  A fraction of the resulting spontaneously emitted light is collected into an optical fiber, with each photon's polarization ($\sigma^+$ or $\sigma^-$) entangled with its parent atom due to atomic selection rules (Fig. \ref{fig:network and microwaves}a). Each photon passes through a quarter-wave plate that maps circular to linear polarization ($\sigma^+ \rightarrow H$ and $\sigma^- \rightarrow V$), and then the two photons interfere on a 50/50 beam-splitter, where detectors monitor the output (see Fig. \ref{fig:network and dream}a and Methods Summary) \cite{matsukevich:2008}.  We select the two-photon Bell states of light $\ket{HV} +e^{i\phi_D}\ket{VH}$, where $\phi_D$ is $0$ or $\pi$ depending on which pair of detectors registers the photons \cite{simon:2003}.  Finally, a series of microwave pulses transfers the atoms into the $\{ \ket{0}, \ket{1} \}$ basis (Fig. \ref{fig:network and microwaves}b), ideally resulting in the heralded entangled state of the two remote atomic qubits $\ket{01} + e^{i\phi_{AB}} \ket{10}$. 

The intermodule phase is given by
\begin{equation}
\phi_{AB} = \phi_D +  \Delta \omega_{AB} t + k c\Delta \tau + k \Delta x  + \Delta\phi_T.
\label{eqn:phase}
\end{equation}
In this equation, the phase evolves with the difference in qubit splittings between module A and B, $\Delta \omega_{AB} = \omega_{0,A}-\omega_{0,B} \approx 2\pi \times 2.5$ kHz, owing to controlled Zeeman shifts \cite{olmschenk:2007}.  The stable geometric phase factors $kc\Delta \tau < 10^{-2}$ and $k \Delta x < 10^{-2}$ result from the difference in excitation time $\Delta \tau < 100$ ps and difference in path length $\Delta x < 3$ cm between each atom and the beam-splitter.  Here $c$ is the speed of light and $k \sim 0.33$ m$^{-1}$ is the wavenumber associated with the energy difference of the photon decay modes (here, the energy difference between $\sigma^+$ and $\sigma^-$ photons).  The final contribution is the stable phase difference of the microwave transfer pulses $\Delta \phi_T$ across the modules.

In previous experiments, entanglement between remote atom spins at rates of 0.002 sec$^{-1}$ was accomplished using atom-photon frequency entanglement \cite{olmschenk:2009}, and at rates of 0.026 sec$^{-1}$ using atom-photon polarization entanglement \cite{matsukevich:2008}.  Here, we dramatically increase the single photon collection efficiency by using high numerical aperture microscope objectives and detecting two out of four Bell states of light emitted by the atoms to achieve a heralded entanglement rate of 4.5 sec$^{-1}$ (see Methods Summary). 

Given a heralded photon coincidence event, we verify entanglement between ion trap modules by measuring atomic state populations and coherences following standard 2-qubit tomography protocols \cite{sackett:2000}. We measure an average entangled Bell state fidelity of $0.78 \pm 0.03$.  Imperfect mode matching at the beam-splitter contributes $0.08 \pm 0.02$ to the infidelity. The measured atom-photon polarization entanglement is 0.92 per ion trap which contributes 0.15 to the remote entangled state infidelity.  We attribute the atom-photon polarization infidelity to spatially inhomogeneous rotations of the photon polarization or polarization-dependent loss. Combining imperfect ion-photon polarization entanglement with imperfect mode matching at the beam-splitter yields an expected fidelity of $0.79 \pm 0.02$, consistent with observation.

\begin{figure*}[tbp]
\includegraphics[width=6.5 in]{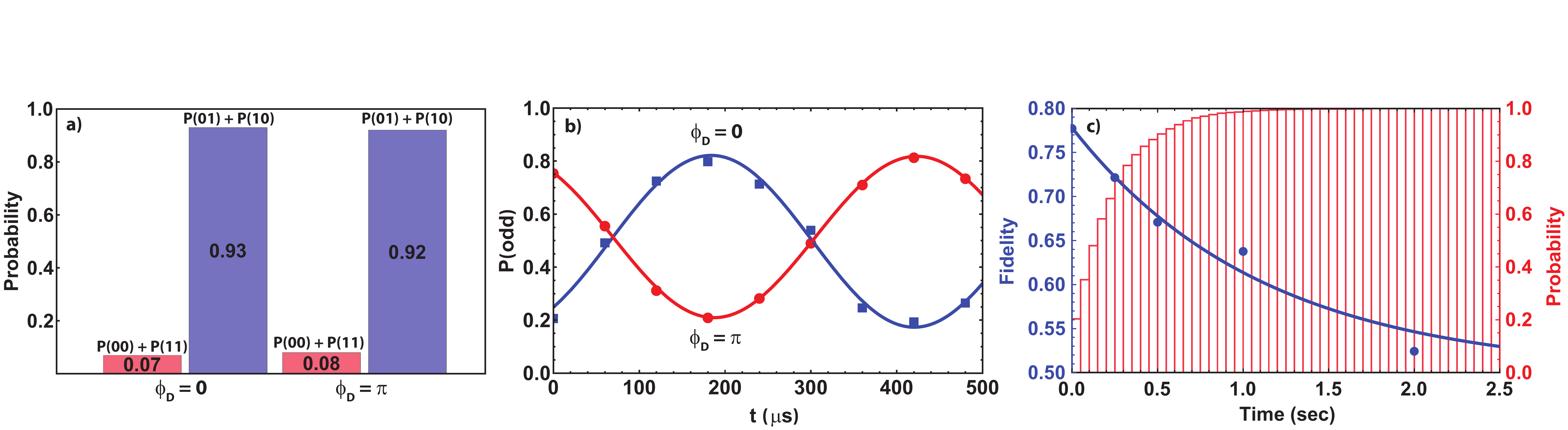}
\caption{\textbf{a)} Populations of two remote atoms after heralding entanglement between modules. After detecting the photon Bell states ($\phi_D =$ 0 or $\pi$), microwave transfer pulses rotate the remote atom populations to the  $\{ \ket{0}, \ket{1} \}$ basis. Subsequent detection of the remotely entangled atoms results in measurement of odd parity, $P(01) + P(10)$, with high probability. \textbf{b)}  Phase coherent time evolution of the remote entangled state with the application of an intermodule magnetic field gradient. After heralding remote entanglement between modules and applying microwave transfer pulses, the addition of a time delay prior to a $\pi/2$ rotation on both atoms results in an out-of-phase oscillatory behavior of the remote atom entangled state with $\phi_D = 0$ or $\pi$ (blue squares and red circles respectively, see Eq. \ref{eqn:phase}). \textbf{c)} Remote entangled state coherence and generation probability vs. time. We measure the remote entangled state coherence time by adding a Ramsey zone delay in the presence of an intermodular magnetic field gradient before application of a spin echo pulse and a $\pi/2$ microwave rotation as described in the text. The decay of the fidelity from the measured loss of phase coherence of the entangled state points to magnetic field gradient noise as the dephasing mechanism. A fit to an exponential function yields a coherence time of 1.12 seconds. The probability of generating entanglement after a given time interval is shown in red. A fit to an exponential function gives the average remote entanglement rate 4.5 sec$^{-1}$.}
\label{fig:lifetime}
\end{figure*}

Since the phase of the entangled state evolves in time (2nd term of Eq. \ref{eqn:phase}), the remote atomic entanglement coherence time can be measured with Ramsey spectroscopy. Unlike a Ramsey experiment with a single atom, this measurement is not sensitive to long-term stability of the local oscillator \cite{olmschenk:2007, chwalla:2007}. We measure the remote entangled state coherence time by repeating the above experiment with constant transfer pulse phase $\Delta \phi_T$ while varying the Ramsey zone delay before a final $\pi/2$ microwave rotation. We utilize a spin echo pulse in the middle of the Ramsey zone delay to account for slow magnetic field gradient drifts, and measure an entanglement coherence time of 1.12 seconds, well in excess of the required time to create remote entanglement between modules (Fig \ref{fig:lifetime}c).

\begin{figure*}[btp]
\includegraphics[width=4.5in]{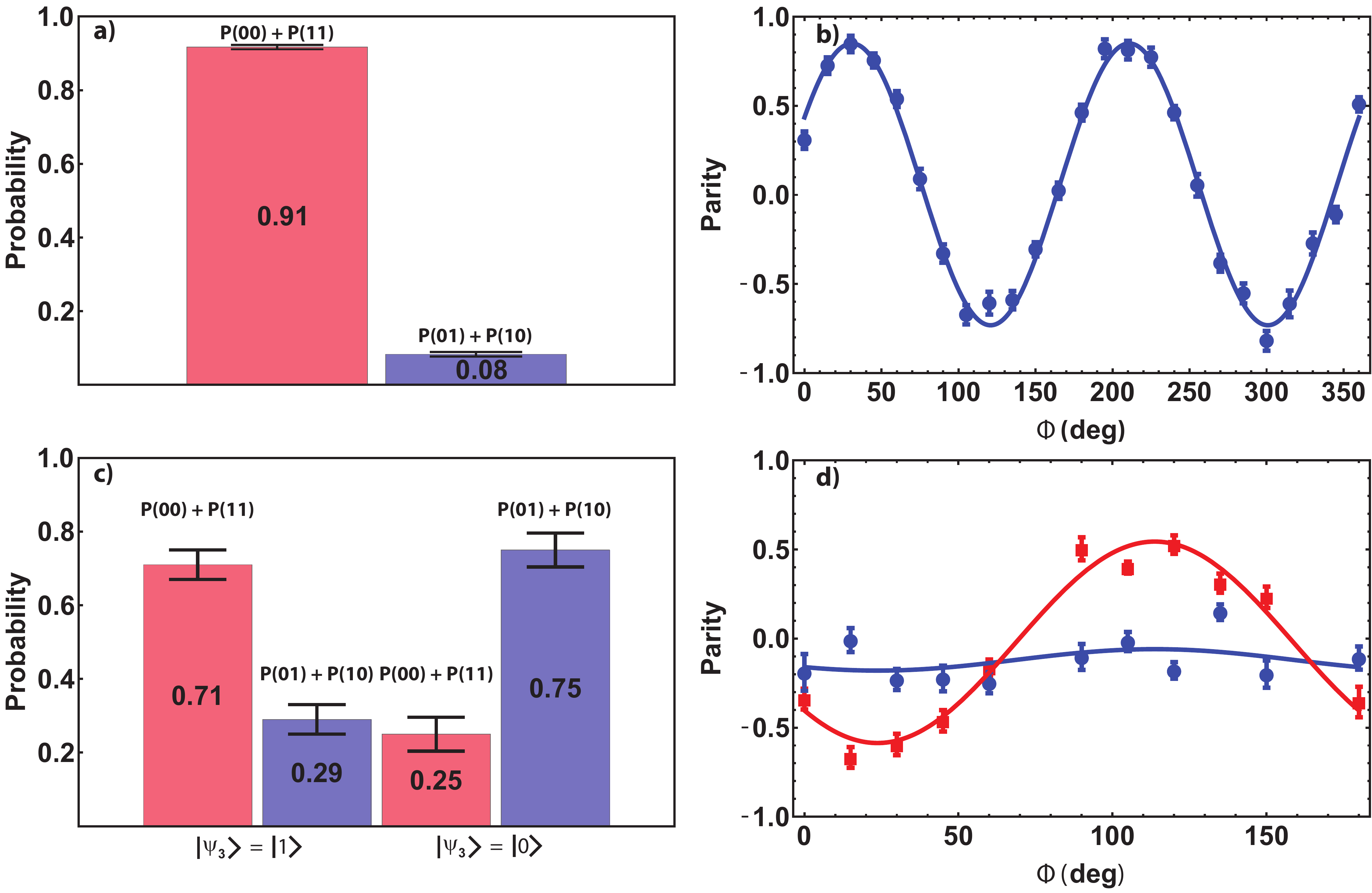}
\caption{ \textbf{a)} Measured populations of two atoms after preparing the atoms in the state $\ket{00}$ and applying an entangling gate through phonons within a module (Eq. \ref{eqn:molmer}), resulting in even parity population $P(00)+P(11) = 0.91 \pm 0.01$, excluding detection error.\textbf{b)} Measured parity of the entangled state following $\pi/2$ qubit rotations with variable phase $\phi$ with respect to intramodular phase $\phi_{A}$ of the two atoms. The amplitude of the parity oscillation is $0.79 \pm 0.02$ and the fidelity of the entangled state is $0.85 \pm 0.01$ excluding state detection errors. \textbf{c)} Populations of two atoms in ion trap module A after remote entanglement between atoms 2 and 3 followed by entanglement between atoms 1 and 2 as described in the text. After measuring the resulting three particle state (see Eq. \ref{eq:tripartite}), if the remote atom is in the state $\ket{1}$, atoms 1 and 2 should be in an even parity state. If the remote atom is in the state $\ket{0}$, atoms 1 and 2 should be in an odd parity state. We observe this correlation with the remote atom with probability $0.71 \pm 0.04$ and $0.75 \pm 0.05$ respectively after averaging over detection of the entangled photon states. \textbf{d)} Parity oscillation of atoms 1 and 2 conditioned on detecting the remote atom in the state $\ket{1}_B$ (red squares) and $\ket{0}_B$ (blue circles). After remote entanglement between modules and entanglement within one module, we apply a Raman $\pi/2$ rotation with variable phase $\phi$ to atoms 1 and 2 in module A and measure the state of all three atoms. If the remote atom is in the state $\ket{0}_B$, a $\pi/2$ rotation on atoms 1 and 2 maps $\ket{\psi_1 \psi_2}_A = \ket{01}_A-i\ket{10}_A$ to a state with zero average parity for any phase $\phi$ of the rotation. If the remote atom is in the state $\ket{1}_B$, a $\pi/2$ rotation with variable phase $\phi$ of $\ket{\psi_1 \psi_2}_A = \ket{00}_A-ie^{-\phi_{A}}\ket{11}_A$ maps the parity of this state to $\cos (\phi_{A}-2\phi)$. We observe such a parity oscillation correlated with the state of the remote atom. The fidelity of the two qubit entangled state $\ket{00}_A - ie^{-i\phi_{A}}\ket{11}_A$ conditioned on detecting the remote atom in $\ket{\psi_3}_B = \ket{1}_B$ is 0.63 $\pm$ 0.03.}
\label{fig:2x1}
\end{figure*}

In addition to using a photonic interconnect between ion traps, we use the Coulomb-coupled transverse phonon modes of the atoms to create entanglement within one module (see Fig. \ref{fig:network and microwaves}c). Off-resonant laser beams drive stimulated Raman transitions between the qubit levels and impart spin-dependent forces detuned from the phonon modes. Following conventional Coulomb gate protocols \cite{molmer:1999, blatt:2008}, after a certain time the motion returns to its original state (see Methods Summary), and the four two-qubit basis states are ideally mapped to the following entangled states
\begin{equation}
\begin{split}
\ket{00} &\rightarrow  \ket{00} -ie^{-i\phi_{A}}\ket{11} \\
\ket{11} &\rightarrow \ket{11} -ie^{i\phi_{A}}\ket{00} \\
\end{split}
\quad \quad
\begin{split}
\ket{01} &\rightarrow  \ket{01}-i\ket{10} \\
\ket{10} &\rightarrow  \ket{10}-i\ket{01},
\end{split}
\label{eqn:molmer}
\end{equation}
where $\phi_{A}$ is the intramodular phase from this optical Raman process in module A \cite{lee:2005}. This phase depends on the relative optical phase of two non-copropagating lasers.
Using the above gate operation on two Doppler-cooled atoms within a module ($\bar{n}\sim3$), we create the state $\ket{00} - ie^{-i\phi_{A}}\ket{11}$ with a fidelity of $0.85 \pm 0.01$, excluding detection error, as shown in Fig. \ref{fig:2x1}a,b.

We now describe the integration of both photonic and phononic  buses to generate entangled 3-particle states. The three atoms are first prepared in the state $\ket{\psi_1 \psi_2}_A \ket{\psi_3}_B = \ket{00}_A\ket{0}_B$ with atoms 1 and 2 in module A and the remote atom 3 in module B (see Fig. \ref{fig:network and dream}a). After heralding entanglement between atom 2 in module A and atom 3 in module B using photons, we re-initialize atom 1 to the state $\ket{0}_A$ with an individual addressing optical pumping beam, and then we entangle atoms 1 and 2 within module A using phonons.  Ideally, this produces the state
\begin{eqnarray}
\ket{\psi_1 \psi_2}_A \ket{\psi_3}_B = \Big ( \ket{00}_{A}-ie^{-i\phi_{A}}\ket{11}_{A} \Big )\ket{1}_{B}  \nonumber \\
+e^{i\phi_{AB}}\Big( \ket{01}_{A} - i\ket{10}_{A} \Big ) \ket{0}_{B}
 \label{eq:tripartite}
\end{eqnarray}
In the above state, the parity of any pair of atoms is correlated with the spin state of the third atom. We take advantage of this property to probe the parity of atoms 1 and 2 in module A, and correlate it with the state of remote atom 3 in module B. After making photon and phonon connections between the atoms, we apply a $\pi/2$ Raman rotation to atoms 1 and 2 with a variable phase $\phi$ followed by state detection of all three atoms. When the remote atom is measured in state $\ket{\psi_3}_B = \ket{1}$, the spin parity of atoms 1 and 2 in module A is $\Pi = \Pi_c \cos (\phi_{A} - 2\phi)$. When the remote atom is measured in state $\ket{\psi_3} = \ket{0}_B$, the atoms in module A should be mapped to a state with zero average parity, regardless of the phase of the $\pi/2$ Raman rotation. We observe this correlation with a remote entangled state generation rate of $\sim$4 sec$^{-1}$ as shown in Fig. \ref{fig:2x1}c,d. The fidelity of detecting the state $\ket{00}_A-ie^{-i\phi_{A}}\ket{11}_A$ of atoms 1 and 2 conditioned on detecting the remote atom 3 in the state $\ket{1}_B$ is $0.63 \pm 0.03$. 

Scaling this architecture to many modules can vastly simplify the complexity of phases to be tracked and controlled.  For $N \gg 1$ modules each with $n \gg 1$ qubits and $m \ll n$ optical ports at each module, the number of overall phases is reduced by a factor of $1/N + (m/n)^2$ compared to that for a fully connected set of $nN$ qubits \cite{musiqcpaper}. Of course in a modular architecture there may be overheads associated the reduced connectivity, but it will be useful to have flexibility in this tradeoff.

The intermodule phase $\phi_{AB}$ in the experiment is easily controlled by setting the phase difference of microwave rotations between the two modules. The intramodule phase $\phi_A$ is determined by the optical phase difference of the two Raman lasers and is passively stable for a single entangling experiment for typical gate times of order 100 $\mu s$. Tracking and controlling the optical phases between many entangled pairs in spatially separated modules at different times can be accomplished by utilizing ``phase insensitive" gates \cite{lee:2005}.

Scaling this system will also require mitigating crosstalk within modules.  For example, when generating photons for intermodular entanglement, laser scatter and radiated light will disturb neighboring qubits within a module.  This may require the use of different species of atoms as photonic and memory qubits. Quantum information could then be transferred from the photonic qubits to the memory qubits via the Coulomb bus \cite{schmidt:2005}.  The second (photonic) species can also be used for intermittent sympathetic cooling \cite{barrett:2003}.

These experiments demonstrate a first step toward a modular architecture using multiple quantum buses to generate entanglement. This modular architecture can be expanded to include many modules and an optical cross connect switch to create a flexible, reconfigurable photonic network between modules (Fig. \ref{fig:network and dream}b) and thus be made fault tolerant for the execution of extended quantum circuits \cite{musiqcpaper}. Modular architectures may be used as the backbone of a quantum repeater network \cite{briegel:1998} and of a quantum network of clocks \cite{komar:2013}. The experiments here suggest a figure of merit for a quantum repeater network with maximum separation between nodes: the coherent entanglement distance $D_{\text{ent}} = d_q R \tau$,  where the physical qubit separation $d_q$ is multiplied by the entanglement rate $R$ and the entangled state coherence time $\tau$. This figure of merit indicates the maximum entanglement distance between modules of a quantum network with a positive output entanglement rate. The experiments presented here give $D_{\text{ent}} = 1$ m $\times$ 4.5 sec$^{-1}$ $\times$ 1.12 sec $\approx$ 5 meters, orders of magnitude larger than previous experiments in any platform. The coherent entanglement distance in this experiment can be lengthened by increasing the remote entanglement rate and entangled state coherence time. In addition, the development of low-loss UV fibers or the efficient down-conversion of photons to telecommunication wavelengths can increase the qubit separation without affecting the entanglement rate and enable long distance quantum repeater networks \cite{pelc:2010}.

\section{Methods Summary}
In this experiment, ion trap module A is a segmented, four blade design useful for holding chains of trapped atoms. A trap drive frequency of 37.15 MHz is used to achieve secular transverse frequencies of $\sim$2.4 MHz. Module B is a four rod Paul trap that confines a single atom. This trap is driven at 37.72 MHz to achieve secular frequencies of $\sim$1.5 MHz.

In order to generate remote entanglement between atoms in physically separated ion traps, we optically pump both atoms to the $\ket{0,0}$ state. A picosecond laser pulse resonant with the ${}^{2}S_{1/2}\rightarrow {}^{2}P_{1/2}$ transition excites trapped atoms in different modules. The atoms spontaneously emit photons of which $\sim$10 \% are collected by a large NA = 0.6 single atom microscope objective, resulting in the entangled photon-polarization, atom-spin state $\frac{1}{2}(\ket{1,1}\ket{\sigma^{-}}- \ket{1,-1}\ket{\sigma^{+}})^{\otimes 2}$. The emitted photons pass through $\lambda/4$ waveplates to convert the photon polarization to linear horizontal (H) or linear vertical (V) resulting in the atom photon state $(\ket{1,1}\ket{V}-i\ket{1,-1}\ket{H})^{\otimes 2}$. Each objective is mode matched to a single-mode optical fiber which delivers the photons to an interferometer with a 50/50 beam-splitter as the central element. The interferometer effects a Bell state measurement of the photon state. We detect two out of the four possible Bell states of light exiting the beam-splitter to herald the entanglement of the remote atoms' spins \cite{simon:2003}, and after a series of microwave transfer pulses, the remote atom entangled state is $\ket{01} + e^{i\phi_{AB}}\ket{10}$ with the intermodule phase $\phi_{AB}$ defined in the main text. The phase $\phi_D$ is 0 if coincident photons are detected on PMTs 1 and 2 or 3 and 4 (see Fig. \ref{fig:network and dream}a). The phase $\phi_D$ is $\pi$ if coincident photons are detected on PMTs 1 and 3 or 2 and 4.

The remote entanglement rate is limited by the collection and detection efficiency of emitted photons from the atoms. The probability for coincident detection of two emitted photons upon exciting both atoms simultaneously with a resonant laser pulse is $P= p_{Bell}[P_{\pi} P_{S_{1/2}} Q_E T_{fib} T_{opt} \frac{\Omega}{4\pi}]^2 = 9.7 \times 10^{-6}$ where $P_{\pi} = 0.95$ is the probability of exciting the atom with a resonant ${}^{2}S_{1/2} \rightarrow {}^{2}P_{1/2}$ laser pulse, ${}^{2}P_{S_{1/2}} = 0.995$ is the probability to decay from ${}^{2}P_{1/2} \rightarrow {}^{2}S_{1/2}$ (as opposed to the ${}^{2}D_{3/2}$ state), $p_{Bell} = 1/2$ accounts for selecting two of the four possible Bell states of light, $Q_E \approx 0.35$ is the quantum efficiency of the single photon PMT detectors, $T_{fib} \approx 0.14$ is the fiber coupling and transmission probability of a single-mode optical fiber, $T_{opt} = 0.95$ is the photon transmission through optical components, and $\frac{\Omega}{4\pi} = 0.1$ is the fraction of the solid angle each microscope objective subtends. The experimental repetition rate of 470 kHz is limited by the need for Doppler cooling (adding $\sim$500 ns on average to the repetition time), the atomic state lifetime of the ${}^{2}P_{1/2}$ state (necessitating $\sim$1 $\mu$s of optical pumping for state preparation of the pure quantum state $\ket{0}$), and sound wave propagation time in AOM crystals used in the experiment. These factors result in a measured atom-atom entanglement rate of 4.5 sec$^{-1}$.

The Coulomb entangling gate makes use of Walsh function modulation $W[1]$ to reduce the sensitivity of the gate to detuning and timing errors \cite{hayes:2012}. We pick a detuning $\delta$ from a transverse mode of motion and set the gate time $t_g = 2 / \delta$ with a $\pi$ phase advance of the sidebands at $t = t_g/2$. We adjust the average Raman laser intensity power to make sideband Rabi frequency $\eta \Omega$ satisfy $\delta = 2^{3/2}\eta \Omega$ to complete the entangling gate $\ket{00} \rightarrow \ket{00}-ie^{-i\phi_{A}}\ket{11}$ in ion trap module A.

Detection error of a single atom in an ion trap module is limited by off-resonant pumping  from the F = 1 to the F = 0 manifold of the ${}^{2}S_{1/2}$ ground state through the F = 1 manifold of the ${}^{2}P_{1/2}$ excited state \cite{olmschenk:2007}, and is $\sim1$\% in the experiments presented here. Detection error of two qubits in the same module is limited by the use of a single PMT detector where the photon detection histograms of a single qubit in the state $\ket{1}$ and two qubits in the state $\ket{11}$ may overlap. This overlap is $\sim8$\% in these experiments.

\section{Acknowledgments}
We thank Kenneth R. Brown, L.-M. Duan, J. Kim, P. Kwiat, D.~N. Matsukevich, P. Maunz,
D.~L. Moehring, S. Olmschenk, and P. Richerme for helpful discussions. This work was supported by the Intelligence Advanced Research Projects Activity, the Army Research Office MURI Program on Hybrid Quantum Optical Circuits, and the NSF Physics Frontier Center at JQI. 
%\section{Author Contributions}
%D.H., I.V.I., G.V., S.M.C., S.D., C.C., and C.M. all contributed to the experimental design, construction, data collection and analysis of this experiment. All authors contributed to this manuscript.
%\section{Competing Financial Interests}
%The authors declare no competing financial interests.

%\bibliographystyle{apsrev4-1}
\bibliographystyle{naturemag}
\bibliography{2x1_2}

\end{document}